\documentclass[ preprintnumbers,preprint,prb]{revtex4}

\usepackage{graphicx}

\newcommand{\ket}[1]{| #1 \rangle}

\def\identity{\leavevmode\hbox{\small1\kern-3.8pt\normalsize1}}

\begin{document}

\centerline {\Large \bf Multi-Qubit Gates in Arrays}

\centerline {\Large \bf Coupled by `Always On' Interactions}

\bigskip

\centerline {\bf Simon C. Benjamin$^*$}

\smallskip

{\footnotesize

\centerline {Ctr. for Quantum Computation, Clarendon Lab., Univ. of Oxford, {\scriptsize OX1 3PU}, UK.}

\centerline {\& Department of Materials, Parks Road, Univ. of Oxford,  {\scriptsize OX1 3PH},  UK.}
\centerline {$^*$ s.benjamin@qubit.org.}

\smallskip
{\bf 
Recently there has been interest in the idea of quantum computing {\em without} control of the physical interactions between component qubits. This is highly appealing since the `switching' of such interactions is a principal difficulty in creating real devices. It has been established that one can employ `always on' interactions in a one-dimensional Heisenberg chain, provided that one can tune the Zeeman energies of the individual (pseudo-)spins. It is important to generalize this scheme to higher dimensional networks, since a real device would probably be of that kind. Such generalisations have been proposed, but only at the severe cost that the efficiency of qubit storage must {\em fall}. Here we propose the use of multi-qubit gates within such higher-dimensional arrays, finding a novel three-qubit gate that can in fact increase the efficiency {\em beyond} the linear model. Thus we are able to propose higher dimensional networks that can constitute a better embodiment of the `always on' concept - a substantial step toward bringing this novel concept to full fruition.}

Quantum computation (QC) holds the promise of manipulating information in fundamentally new ways, performing tasks that are effectively impossible with any classical computer. However, the challenge of experimentally realizing such a device is extremely difficult. 
The archetypal theoretical model for QC assumes that one can freely perform various kinds of manipulation on the elementary qubits. (Here we will use the term `qubit' to refer to the units of quantum information, and the term `spin' to refer to the physical two-state systems that represent the qubits - an actual electron spin being one obvious example.) The required manipulations include all possible transformations on each qubit in isolation, the so-called one-qubit gates, together with the ability to switch `on' and `off' the interactions between spins in order to perform an entangling two-qubit gate. A literal embodiment of this model would therefore call for at least two different forms of control, which together can achieve arbitrary local manipulations of this kind\cite{notNMR}. There do exist many such schemes, for example the Kane proposal\cite{kane} where a combination of global EM pulses and local eletrostatic gates is required. However, there have been several theoretical efforts to derive alternatives to the archetypal model, which would be less difficult to implement experimentally. Considerable attention has been focused on the idea of `all Heisenberg' computation\cite{3qubitExchangeOnly, Levy, myABqubitPaper}, where one only needs to manipulate the interactions between spins (dispensing with the need to separately manipulate isolated spins), at the cost of encoding each qubit over two or three spins. This idea is clearly a powerful simplification, however in many systems it is in fact the control of the spin-spin interactions that is the most difficult aspect. A related idea is that of global control: here one dispenses with the ability to localize manipulations and instead applies all manipulations globally across the entire device\cite{ababPRL}. Again, this idea still requires one to switch interactions, albeit collectively (unless they are of the relatively passive Ising type \cite{oldPRA}). 

Recently however, ideas have been advanced for performing quantum computation in strongly coupled systems {\em without} the need to switch interactions \cite{ourPRL, zhou, newLANL, newLongPaper}. It has been shown that one can perform all the necessary one-qubit and two-qubit gates simply by `tuning' the Zeeman splitting of the individual qubits. 
This approach involves using some spins to hold qubits, while others are employed to act as passive barriers between qubits. Interestingly, this `barrier-spin' concept proves to be compatible with the concept of global control, so that in one variant of the idea it is not necessary to `target' the Zeeman tuning onto single spins within the device\cite{ourPRL}. There is a cost associated with all existing variants of the scheme: at least one barrier spin is required for each qubit-bearing spin, and therefore the ratio $R$ of qubits to spins is limited by $R=1/2$. This ratio is one important measure of the efficiency of an architecture, and frequently the QC schemes that aim to lighten experimental requirements come with the cost that $R$ must be significantly less than unity. For example, the `all Heisenberg' scheme mentioned above requires $R=1/3$ (assuming an isotropic environment).
However, the barrier scheme has so far only been fully explored for the case of a one-dimensional system. This is a significant limitation: although in principle quantum algorithms can be performed on one-dimensional architectures without losing their advantage over classical machines, in practice two- or three-dimensional structures will prove considerably more effective\cite{1Dlimits}. There has been a brief discussion of the generalization of the 1D results to higher dimensional networks\cite{newLongPaper}, however the approach taken there comes at the cost that the proportion of inactive barrier spins must {\em increase} and consequently $R$ falls to $R=2/5$ at best. It would therefore be very desirable to find a new approach to implementing 2D and 3D barrier networks, whereby the efficiency can at least match the 1D case, if not exceed it. This is the problem we address in the present paper. 

We will introduce the necessary concepts by referring to the earlier work on one-dimensional arrays, and then continue to examine the multi-qubit nodes that are possible in higher dimensional networks, and which (we will show) can offer superior values of $R$. We will present the analysis through an explicit matrix diagonalisation method that is, we hope, straightforward to comprehend for the interested non-specialist. 
In our models we will assume a general anisotropic Heisenberg interaction of the form $J_Z\sigma^Z_1\sigma^Z_2+J_{XY}(\sigma^X_1\sigma^X_2+\sigma^Y_1\sigma^Y_2)$, which includes the well known isotropic form, and the purely planar ``XY'' interaction, as special cases. There is a considerable range of promising physical systems associated with this spectrum of interactions (for the isotropic limit, see e.g. Refs. [\onlinecite{DiVincenzo1,kane,spinResTrans}], and for the anisotropic case, Refs. [\onlinecite{Imamoglu,Mozyrsky,Seiwert}]). Note that the XY limit is also relevant to F\"orster-Dexter processes, e.g. in the context of excitonic exchange in biological molecules.

\begin{figure}[!t]
\centering
\resizebox{7.3cm}{!}{\includegraphics{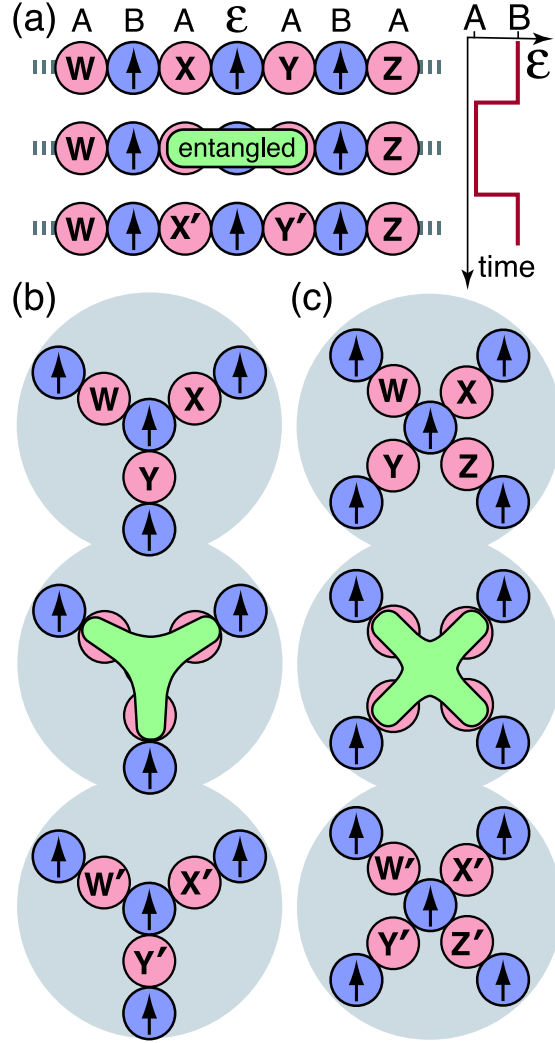}}

\vspace{0.2cm}

\caption{Possible geometries for the `barrier' QC scheme. (a) The established one-dimensional scheme. Letters $A$, $B$, {\Large $\epsilon$} denote Zeeman energies; $A$ and $B$ are fixed but {\Large $\epsilon$} is abruptly changed as shown by the graph to the right. Essentially the barrier is lowered (brought into resonance) to allow the neighboring qubits to interact, and is subsequently raised later when the barrier `revives' into a product state. The lower illustrations show the logical extension to three-qubit (b) and four-qubit (c) nodes, which can be supported in higher dimensional arrays. Successive illustrations (vertically) correspond to the three rows in the 1D case above. Fig. 2 shows how these nodes might be combined to form networks. }

\label{figure1}

\end{figure}

\bigskip
\noindent {\bf The Heisenberg-to-Ising Transition and the Barrier Scheme.}

Consider a chain of spins coupled via the Heisenberg interaction. Suppose that each coupled pair is far off-resonance, in the sense that the difference between their Zeeman energies is much greater than the coupling strength $J_{XY}$. In this limit a native Heisenberg interaction $J_Z\sigma^Z_1\sigma^Z_2+J_{XY}(\sigma^X_1\sigma^X_2+\sigma^Y_1\sigma^Y_2)$ tends to an {\em effective} Ising coupling $J_Z\sigma^Z\sigma^Z$ (for a discussion of this phenomenon, see Ref. \onlinecite{newLongPaper}). This can be exploited as the basic mechanism for quantum gate operations, if one assumes that the Zeeman discrepancy can be dynamically {\em tuned} \cite{ourPRL}: one would tune a set of spins into resonance to perform the gate, and then restore the discrepancy to return the device to its passive state. Analysis of the process is simplified if we assume very abrupt tuning, but this is not a necessary condition \cite{newLongPaper}. In Ref. [\onlinecite{ourPRL}] we showed how to perform this process on a one-dimensional chain. We introduced the idea of using a `barrier' spin between the qubit-bearing spins, such that the barrier becomes entangled with the qubits only during gate operations (see Fig. 1, upper part). This barrier spin is initially in a known eigenstate, $\ket{\uparrow}$ say, and the duration of the on-resonance phase must be such that the barrier returns to this eigenstate. This condition corresponds to achieving coincidence in the revival times of different computational basis states, as will be seen presently. Previous work has only shown that this is possible for barriers separating {\em two} qubits. This is of course the only case of interest for a one-dimensional chain (given nearest-neighbor interactions), however it is not the natural arrangement for two- and three-dimensional networks. 

In principle one {\em can} generate a barrier-based network by taking {\em any} arrangement of qubit-bearing spins, and simply introducing a barrier-spin between each interacting pair. However, a moment's reflection reveals that this inevitably leads to a rather large proportion of `wasted' barrier-spins. Recall $R$ is the ratio of qubit-bearing spins to the total number of spins in the device. Clearly, one would wish $R$ to be as high as possible, given the difficulty of protecting spins from decoherence etc. Let us assume here that each qubit is represented by the state of a single spin (this does not preclude encoding of `logical' qubits over several of these elementary qubits, for the purpose of fault tolerance etc). Then for a one-dimensional array, every other spin is acting as a barrier and therefore $R=1/2$. For a two- or three-dimensional array, it must be the case that at least some of the qubits are capable of interacting with more than two neighbors (or else we are isomorphic to the 1D case).  But for each such `additional' interaction, we must have a barrier spin to control it. Thus $R<1/2$ for all arrays using this architecture. If we restrict ourselves to considering simple regular arrays where each qubit interacts with exactly $n$ neighboring qubits, then we readily conclude that the optimal value $R=2/5$ corresponds to $n=3$. (Notice that if we inspect Fig. 2 (b) and (c), but assign the opposite colour scheme to that specified in the Figure, i.e. make red$\rightarrow$barrier \& blue$\rightarrow$qubit,  then these two geometries achieve this limit).

It seems regrettable that moving to a higher dimensional array, with the attendant advantages in terms of connectivity, parallelism, etc, must apparently necessitate a lower value of $R$. But in fact this is only true if we limit ourselves to considering networks in which there is one barrier spin for {\em each} adjacent pair of qubits. If we can allow a barrier spin to do `double duty', in the sense that it controls more than one interaction, then we can achieve higher values. For example, consider again Fig.2(b) \& (c) but now take the red colour as denoting a qubit, and the blue as denoting a barrier. The value of $R$ thus increases from $2/5$ to $3/5$ and we have an architecture that is {\em more} efficient than the one-dimensional case. Similarly, the geometry shown in (d) would offer $R=2/3$. However, the difficulty is that `lowering' such a barrier will lead to a much more difficult condition in terms of the state revival: the barrier must, at some subsequent time, be dis-entangled from all adjacent qubits (i.e., must return to a product state) in order that it can be `raised' again. We will now consider whether this condition can be met for the two simplest cases: the three-qubit mode, as shown in Fig.1(b), and the four qubit node illustrated in Fig. 1(c).

\begin{figure}[!t]
\centering
\resizebox{10.7cm}{!}{\includegraphics{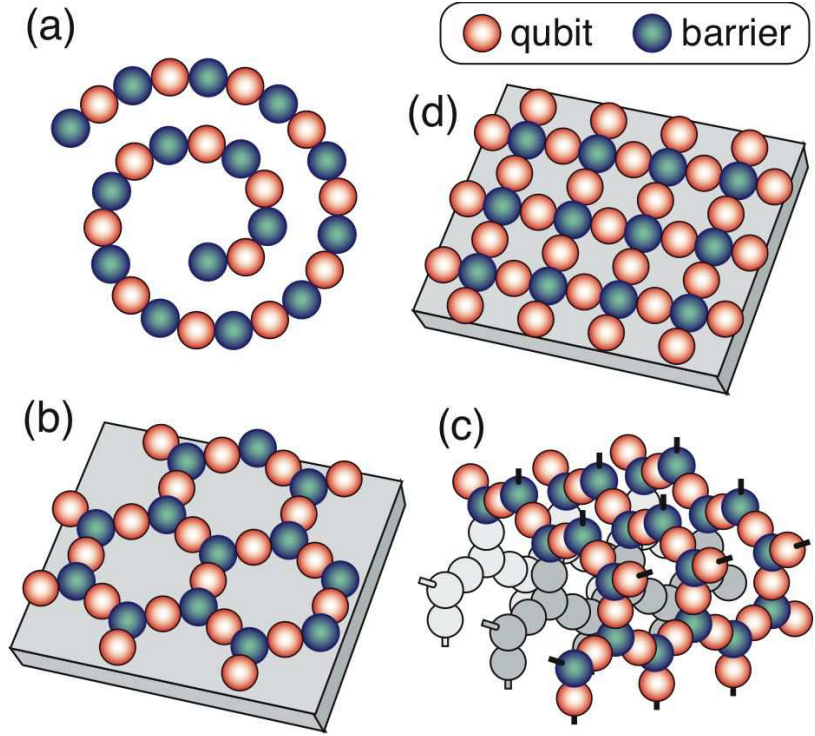}}

\vspace{0.2cm}

\label{figure2}
\caption{Illustration of the geometries that can be formed from the multi-qubit gates depicted in Fig. 1. Here (a) is the simple $ABAB...$ linear array that is established in the literature. Parts (b) \& (c) show examples of the two- and three-dimensional arrays that can be generated purely from a node of the kind shown in Fig. 1(b). (Fortunately, hexagonal patterns are in fact quite common occurrences in self-assembling nanostructures.) In contrast to the networks suggested previously, here the nodes of the structure are barrier spins, and therefore each such barrier separates more than two qubits. In (d) we see one example of the structure that could be formed from the four-qubit node depicted in Fig. 1(d).}

\end{figure}

\bigskip
\noindent{\bf The Three Qubit Node.}

Consider a sub-network of seven spins with the topology and the initial state shown in Fig. 1(b), top. This can be seen as a `fragment' of either the 2D pattern of Fig. 2(b) or the 3D pattern in Fig 2(c). There are four barrier spins in the eigenstate $\ket{\uparrow}$ and three qubit bearing spins, the qubits being labelled $W$, $X$ and $Y$.  Initially we suppose the device is in a `passive' state. The physical distinction between the barrier spins and the qubit-bearing spins is simply that the two types have far off-resonance Zeeman energies, so that the underlying Heisenberg interaction between a barrier spin and a qubit spin is effectively of the Ising form. Thus the barriers will remain in their $\ket{\uparrow}$ eigenstates indefinitely (or more accurately, for a finite time depending on the accuracy of the Ising approximation, which depends \cite{newLongPaper} on the magnitude of the ratio of the detuning to the interaction strength $|A-B|/J$). 

Now suppose that we abruptly tune the Zeeman energy of the central barrier spin so that it is comparable to the Zeeman energy of the qubit-bearing spins. (The transition need not be abrupt\cite{newLongPaper}, but this assumption vastly simplifies the analysis). Notice that the three outer barrier spins, being still far off-resonance from the qubit spins, will continue to interact via an Ising form $J\sigma_Z\sigma_Z$ - and moreover they will remain in the state $\ket{\uparrow}$ throughout.  Therefore their contribution to the Hamiltonian of the central four spins is simply a shift to the Zeeman energy of their neighbors. Let us label the three spins which initially bear qubits by the numbers $1$, $2$, $3$, and the middle spin by the letter $M$. Then we can write the Hamiltonian of these spins as 

$$
H_{\rm quadruplet}=H_{\rm zeeman}+H_{\rm int}
$$
$$
H_{\rm zeeman}=B\sigma_M^Z+(A+ J_Z)\sum_{j=1}^3\sigma_j^Z
$$
\begin{equation}
H_{\rm int}=\sum_{j=1}^3J_{XY}(\sigma_j^X\sigma_M^X+\sigma_j^Y\sigma_M^Y)+J_Z\sigma_j^Z\sigma_M^Z
\label{hamil}
\end{equation}
It will prove interesting to see the consequences of planar anisotropy ($J_Z\neq J_{XY}$) when we attempt to achieve a complete gate. A full representation of this Hamiltonian requires a matrix with $16=2^4$ rows/columns. However it is of course block-diagonal, as illustrated in Fig. 3(a), because $H_{int}$ commutes with $S_Z=\sigma_M^Z+\sum \sigma^Z_j$. Thus we can understand the complete dynamics by considering the motion within each subspace separately. In the figure we introduce the notation $H_{X,Y}$ for the Hamiltonian within a given subspace, where the subscript $(X,Y)$ indicates the space is spanned states having $X$ `up' spins and $Y$ `down' spins (equivalently, one could of course label by the $S_Z$ eigenvalues). We will establish the eigenstructure of these subspaces, and then consider whether the `revivals' in each space can be made to coincide.

\begin{figure}[!t]
\centering
\resizebox{7.7cm}{!}{\includegraphics{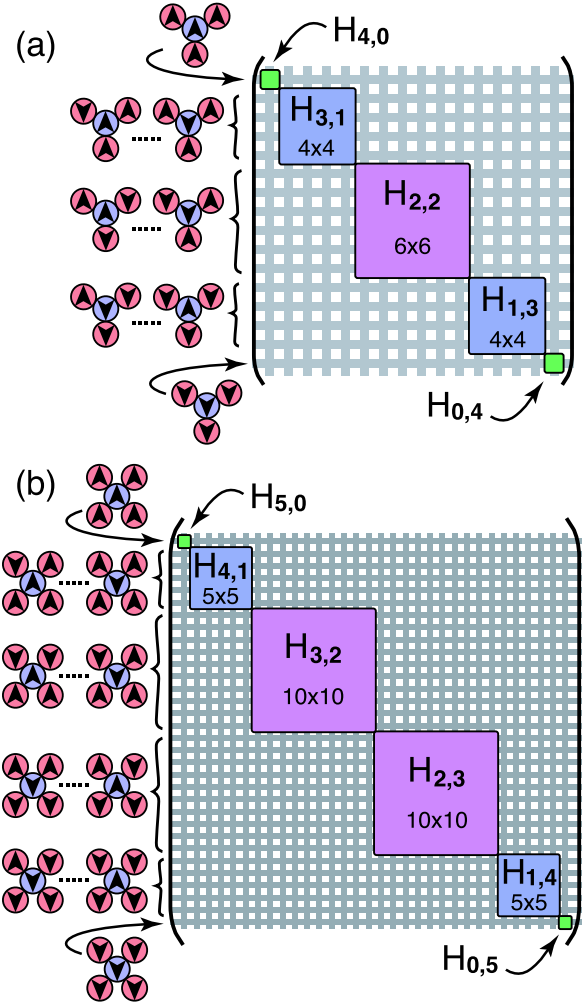}}

\vspace{0.2cm}

\caption{Diagram showing the block-diagonal structure of the Hamiltonian for the three-qubit node (a) and the four-qubit node (b). The spin basis states are shown schematically: the notation $\ket{\uparrow\uparrow\uparrow}\ket{\uparrow}$, for example, corresponds to the uppermost row in part (a).}

\label{figure3}

\end{figure}

For the complete set of 16 spin basis states, we use a notation of the form $\ket{\downarrow\downarrow\downarrow}\ket{\uparrow}$ where the right ket contains the state of the central spin, and the left ket contains the states of the outer three spins, in clockwise order from the top left (although in fact the order will not matter to us). If we write the initial computational basis states as $\ket{000}$, $\ket{001}$...$\ket{111}$, with the same ordering, then recalling that the initial barrier state is $\ket{\uparrow}$ we simply identify these initial states with the eight of the sixteen spin states thus:
\begin{equation}
\begin{array}{ccc}
 \ket{000} & \rightarrow & \ket{\downarrow\downarrow\downarrow}\ket{\uparrow} \\
 \ket{001} & \rightarrow & \ket{\downarrow\downarrow\uparrow}\ket{\uparrow} \\
&\vdots& \\
 \ket{111} & \rightarrow & \ket{\uparrow\uparrow\uparrow}\ket{\uparrow}
\end{array}
\label{mapping}
\end{equation}

Firstly we note that there are two trivial subspaces associated with the states $\ket{\uparrow\uparrow\uparrow}\ket{\uparrow}$ and $\ket{\downarrow\downarrow\downarrow}\ket{\downarrow}$ (though the latter is not of interest). The computational basis state $\ket{111}$ therefore remains an eigenstate with energy $3(A+J_Z)+B+3J_Z$.
We introduce the symbols $a\equiv A+J_Z$ and $\delta\equiv B-a$, so that this energy is $4a+\delta+3J_Z$.

Now the smallest non-trivial subspaces are those with Hamiltonians $H_{3,1}$ and $H_{1,3}$. As required by symmetry, these Hamiltonians prove to have the same structure:
\[
H=K\identity + 2J_{XY}\left( 
\begin{array}{cccc}
0 & 0 & 0 & 1 \\ 
0 & 0 &  0 & 1 \\ 
0 & 0 &  0 & 1\\ 
1 & 1 &  1 & X
\end{array}
\right)
\]
where $\identity$ is the identity matrix. For the $H_{3,1}$ subspace $K=K_{3,1}=2a+\delta+J_Z$, $X=X_{3,1}=-(\delta+2J_Z)/J_{XY}$, and the basis runs $\{\ket{\downarrow\uparrow\uparrow}\ket{\uparrow}$, $\ket{\uparrow\downarrow\uparrow}\ket{\uparrow}$, $\ket{\uparrow\uparrow\downarrow}\ket{\uparrow}$, $\ket{\uparrow\uparrow\uparrow}\ket{\downarrow}\}$. Notice that the fourth basis state is not a possible initial computational basis state. For the complimentary $H_{1,3}$ subspace we have $K_{1,3}=-2a-\delta+J_Z$ and $X_{1,3}=(\delta-2J_Z)/J_{XY}$, the basis running $\{\ket{\uparrow\downarrow\downarrow}\ket{\downarrow}$, $\ket{\downarrow\uparrow\downarrow}\ket{\downarrow}$, $\ket{\downarrow\downarrow\uparrow}\ket{\downarrow}$, $\ket{\downarrow\downarrow\downarrow}\ket{\uparrow}\}$. Notice that this time {\em only} the fourth basis state is a possible initial computational basis state. 

The eigenstates/values in the two spaces of course have the same form. We may write the (unnormalised) eigenvectors compactly as:

\[
\Bigg\{ \ket{\rm{asym}_1}=\left( 
\begin{array}{c}
1  \\ 
0  \\ 
-1  \\ 
0 
\end{array}
\right) ,\ \ \ket{\rm{asym}_2}=\left(
\begin{array}{c}
1  \\ 
-2  \\ 
1   \\ 
0 
\end{array}
\right)
\Bigg\} \ \ \ \ \ \ \  \
\ket{V_-}=\left( 
\begin{array}{c}
1  \\ 
1  \\ 
1  \\ 
V_- 
\end{array}
\right) \ \ \ \ \ \ \  \
\ket{V_+}=\left( 
\begin{array}{c}
1  \\ 
1  \\ 
1  \\ 
V_+ 
\end{array}
\right)
\]
Vectors $\ket{V_\pm}$ have eigenvalues $E=V_\pm=\frac{1}{2}\left(X\pm(12+X^2)^{1/2}\right)$, while the paired vectors are degenerate with value $E=0$. Here $X$ is of course either $X_{3,1}$ or $X_{1,3}$ for the two respective subspaces. From these eigenvalues, the full energies are obtained as $K + 2J_{XY}E$.

We are now in a position to examine how certain qubit basis states evolve. Consider the basis state $\ket{000}$. This corresponds to $\ket{\downarrow\downarrow\downarrow}\ket{\uparrow}$ and is therefore in the subspace of $H_{1,3}$.
\[
\ket{000}=
\left(
\begin{array}{c}
0 \\
0 \\
0 \\
1 
\end{array}
\right)
=
N(V_+\ket{V_-}-V_-\ket{V_+})
\]
At time $t$ after the Zeeman shift event, we will have:
\[
\ket{000}\rightarrow N^\prime\exp(-i\tau(-2a-\delta+J_Z+2J_{XY}V_-))\Big(V_+\ket{V_-}-V_-\exp(-i\tau2J_{XY}(V_+-V-))\ket{V_+}\Big)
\]
where we have used $\tau\equiv t/\hbar$. Consider the phase difference between $\ket{V_-}$ and $\ket{V_+}$. If this is anything other than the original $-1$, then the state will some finite projection onto $\{\ket{\uparrow\downarrow\downarrow}\ket{\downarrow}$, $\ket{\downarrow\uparrow\downarrow}\ket{\downarrow}$, $\ket{\downarrow\downarrow\uparrow}\ket{\downarrow}\}$. Since these states are not computational basis states, we immediately have our revival condition: $\tau_R$ must be such that $\exp(-i\tau2J_{XY}(V_+-V-))=+1$. This yields $\tau_R=m\pi(12J_{XY}^2+(\delta-2J_Z)^2)^{-1/2}$ for an integer $m$, as our first revival time criterion. At such times we can abruptly shift the Zeeman energy back to its large detuning, and thus decouple the barrier spin. If we do so, the net effect will be simply to introduce a phase: $\ket{000}\rightarrow\exp(i\phi)\ket{000}$ where we find $\phi=\tau_R(2a+\delta-J_Z-2J_{XY}V_-)$. Note that this phase in fact differs from the passive phase $\phi_{\rm passive}=\tau(2a-\delta+3J_Z)$ that {\em would be} accumulated over the same period by a state $\ket{000}$ {\em if there had been no $J_{XY}$ interaction}. This phase difference must be remembered when we come to describe the process as an {\em effective} operation within the space of qubit basis states.

Following analogous arguments we can examine the evolution for the basis states $\{\ket{011},\ket{101},\ket{110}\}$ which are governed by $H_{3,1}$. Now the revival condition is that any state within the space spanned by this triplet must return to it; or equivalently, there must be no projection onto the fourth, non-computational basis state $\ket{\uparrow\uparrow\uparrow}\ket{\downarrow}$. In general a state $\ket{s}$, which at $t=0$ is orthogonal to $\ket{\uparrow\uparrow\uparrow}\ket{\downarrow}$,  will develop as:
\[
\ket{s}\rightarrow 
N\exp(i\phi_1)
\Big(
\big(\alpha\ket{{\rm asym}_1}+\beta\ket{{\rm asym}_2}\big)+
\exp(i\phi_2)\big(V_+\ket{V_-}-V_-\exp(i\phi_3)\ket{V_+}\big)
\Big)
\]
We find that our condition of zero projection onto $\ket{\uparrow\uparrow\uparrow}\ket{\downarrow}$ corresponds to the constraint $\exp(i\phi_3)=+1$, which implies $\tau=n\pi(12J_{XY}^2+(\delta+2J_Z)^2)^{-1/2}$. This revival condition must be met simultaneously with the former condition, and indeed with the further conditions we are about to derive. 

The $H_{2,2}$ subspace remains to be characterized. We find:
\[
H_{2,2}= (\delta-J_Z)\ \identity + 2J_{XY}\left( 
\begin{array}{cccccc}
0 & 0 & 0 & 1 & 1 & 0 \\ 
0 & 0 & 0 & 1 & 0 & 1 \\ 
0 & 0 & 0 & 0 & 1 & 1 \\ 
1 & 1 & 0 & x & 0 & 0 \\ 
1 & 0 & 1 & 0 & x & 0 \\ 
0 & 1 & 1 & 0 & 0 & x 
\end{array}
\right)
\]
Here $x=-\delta/J_{XY}$ and our basis is $\{\ket{\uparrow\downarrow\downarrow}\ket{\uparrow}$, $\ket{\downarrow\uparrow\downarrow}\ket{\uparrow}$,  $\ket{\downarrow\downarrow\uparrow}\ket{\uparrow}$, $\ket{\uparrow\uparrow\downarrow}\ket{\downarrow}$, $\ket{\uparrow\downarrow\uparrow}\ket{\downarrow}$, $\ket{\downarrow\uparrow\uparrow}\ket{\downarrow}\}$. The eigenstates include two degenerate pairs: 
\[
\Bigg\{ 
\ket{V_{4+}a}=\left( 
\begin{array}{c}
1  \\ 
0  \\ 
-1  \\ 
V_{4+} \\
0 \\
-V_{4+} 
\end{array}
\right) ,\ \  \ket{V_{4+}b}=\left( 
\begin{array}{c}
1  \\ 
-2  \\ 
1  \\ 
-V_{4+} \\
2V_{4+} \\
-V_{4+} 
\end{array}
\right) 
\Bigg\} \ \ \ \ \ \ \ \& \ \ \ \
\Bigg\{ 
\ket{V_{4-}a}=\left( 
\begin{array}{c}
1  \\ 
0  \\ 
-1  \\ 
V_{4-} \\
0 \\
-V_{4-} 
\end{array}
\right) ,\ \  \ket{V_{4-}b}=\left( 
\begin{array}{c}
1  \\ 
-2  \\ 
1  \\ 
-V_{4-} \\
2V_{4-} \\
-V_{4-} 
\end{array}
\right)
\Bigg\}
\]
with eigenvalues $V_{4\pm}=\frac{1}{2}\left(x\pm(4+x^2)^{1/2}\right)$. The remaining, non-degenerate eigenvectors are 
\[
 \ket{V_{16+}}=\left( 
\begin{array}{c}
2  \\ 
2  \\ 
2  \\ 
V_{16+} \\
V_{16+}  \\
V_{16+}  
\end{array}
\right)
\ \ \ \ \& \ \ \ \ \
 \ket{V_{16-}}=\left( 
\begin{array}{c}
2  \\ 
2  \\ 
2  \\ 
V_{16-} \\
V_{16-}  \\
V_{16-}  
\end{array}
\right)
\]
with eigenvalues $V_{16\pm}=\frac{1}{2}\left(x\pm(16+x^2)^{1/2}\right)$. These eigenvalues $E=\{V_{4\pm},\ V_{16\pm}\}$ of course correspond to total energies $\delta-J_Z+2J_{XY}E$.

Having thus determined the eigenstructure of this subspace, we can again consider the evolution of the corresponding computational basis states: $\{\ket{100}$,$\ket{010}$,$\ket{001}\}$. We find that any state initially within the space spanned by this triplet will develop as:

\[
\begin{array}{c}
\exp(i\phi_1)
\Bigg(A\Big(V_{4+}\big(\alpha\ket{V_{4-}a} + \beta\ket{V_{4-}b}\big)-
\exp(i\phi_2)V_{4-}\big(\alpha\ket{V_{4+}a} + \beta\ket{V_{4+}b}\big)\Big)\\
\ \ \ \ \ \ \ \ \ \ \ \ \ \ \ \ \ \ \ \ \ \ \ \ \ \ \ \ \ \ \ \ \ \ \ \ \ \ \ \ \ \ \ \ \ \ +B\exp(i\phi_3)\Big(V_{16+}\ket{V_{16-}} + \exp(i\phi_4)V_{16-}\ket{V_{16+}b}\Big)
\Bigg)
\end{array}
\]
\noindent (where $A$, $B$, $\alpha$, $\beta$ are constants depending on the initial state).

In order for this state to be within the original space we must meet two phase conditions:
$\exp(i\phi_2)=+1$ and $\exp(i\phi_4)=+1$. These correspond to two constraints on the revival time:
$\tau_R=p\pi(4J_{XY}^2+\delta^2)^{-1/2}$ and $\tau_R=q\pi(16J_{XY}^2+\delta^2)^{-1/2}$. Now we can collect the four conditions on $\tau_R$ and look for a common solution:
\[
\frac{\tau_R}{\pi}=
\left\{
\begin{array}{c}
m(12J_{XY}^2+(\delta-2J_Z)^2)^{-1/2}\\
n(12J_{XY}^2+(\delta+2J_Z)^2)^{-1/2}\\
p(4J_{XY}^2+\delta^2)^{-1/2}\\
q(16J_{XY}^2+\delta^2)^{-1/2}
\end{array}
\right.
\]

\smallskip
\noindent{\bf The Limit of a Pure XY Interaction (F\"orster Type Interaction).}

Consider the limit where $J_Z=0$. Then the first two constraints are equivalent, and we can rewrite the conditions as
\[
\frac{J_{XY}}{\pi }\tau_R=
\left\{
\begin{array}{c}
n(12+d^2)^{-1/2}\\
p(4+d^2)^{-1/2}\\
q(16+d^2)^{-1/2}
\end{array}
\right.
\]
where we have taken $J_{XY}$ to the right and introduced $d\equiv\delta/J_{XY}$.
In general, there is no value of $d$ for which one can find integers $n$, $m$, $q$ to satisfy the conditions. We can find values of $d$ such that two of the three roots $()^{-1/2}$ are compatible (i.e. in a rational ratio) but the third will not be. For example, $d=0$ makes the second two compatible via $p=2q$, but the first then differs by an irrational factor $\sqrt{3}$. Thus we conclude that our gate cannot operate in the strict limit $J_Z=0$.

\bigskip
\noindent{\bf The Limit of a Pure Isotropic Interaction.}

Consider instead the case where $J_Z=J_{XY}\equiv J$. In this case there is a value of $\delta$ for which all four conditions can be met:  namely the trivial value $\delta=0$. Then:
\[
\frac{\tau_R}{\pi}=
\left\{
\begin{array}{c}
m/(4J)\\
n/(4J)\\
p/(2J)\\
q/(4J)
\end{array}
\right.
\]
and the first revival will be $\tau_R=\pi /(2J)$ implying $m=n=q=2$ and $p=1$. Given this value of $\tau$ we can explicitly write out the transformation in the computational basis. We find: 
\begin{equation}
U=\left( 
\begin{array}{cccccccc}
-1 & 0 & 0 & 0 & 0 & 0 & 0 & 0 \\
\ 0 & -1/3 & 2/3 & 2/3 & 0 & 0 & 0 & \ 0 \\
\ 0 & 2/3 & -1/3 & 2/3 & 0 & 0 & 0 & \ 0 \\
\ 0 & 2/3 & 2/3 & -1/3 & 0 & 0 & 0 & \ 0 \\
\ 0 & 0 & 0 & 0 & 1/3 & -2/3 & -2/3 & \ 0 \\
\ 0 & 0 & 0 & 0 & -2/3 & 1/3 & -2/3 & \ 0 \\
\ 0 & 0 & 0 & 0 & -2/3 & -2/3 & 1/3 & \ 0 \\
\ 0 & 0 & 0 & 0 & 0 & 0 & 0 & \ 1 
\end{array}
\right)
\label{Umat}
\end{equation}
The basis order here is $\{\ket{000}$, $\ket{001}$, $\ket{010}$, $\ket{100}$, $\ket{110}$, $\ket{101}$, $\ket{011}$, $\ket{111}\}$.
It is interesting to note that $U^2=\identity$, thus the second revival at $2\tau_R$ simply yields the indentity: a complete return to the initial state. However $U$ itself is an entangling operator and one can generate more familiar operations by combining it with single qubit operations. For example, one can\cite{composites} obtain:
\[
S=\left( 
\begin{array}{cccccccc}
1 & 0 & 0 & 0 & 0 & 0 & 0 & 0 \\
0 & 1 & 0 & 0 & 0 & 0 & 0 & 0 \\
0 & 0 & 0 & 1 & 0 & 0 & 0 & 0 \\
0 & 0 & 1 & 0 & 0 & 0 & 0 & 0 \\
0 & 0 & 0 & 0 & 1 & 0 & 0 & 0 \\
0 & 0 & 0 & 0 & 0 & 0 & 1 & 0 \\
0 & 0 & 0 & 0 & 0 & 1 & 0 & 0 \\
0 & 0 & 0 & 0 & 0 & 0 & 0 & 1
\end{array}
\right)
\ \ \ \ \ \ \ \ \ 	
P=\left( 
\begin{array}{cccccccc}
-1 & 0 & 0 & 0 & 0 & 0 & 0 & 0 \\
0 & 1 & 0 & 0 & 0 & 0 & 0 & 0 \\
0 & 0 & 1 & 0 & 0 & 0 & 0 & 0 \\
0 & 0 & 0 & 1 & 0 & 0 & 0 & 0 \\
0 & 0 & 0 & 0 & 1 & 0 & 0 & 0 \\
0 & 0 & 0 & 0 & 0 & 1 & 0 & 0 \\
0 & 0 & 0 & 0 & 0 & 0 & 1 & 0 \\
0 & 0 & 0 & 0 & 0 & 0 & 0 & -1
\end{array}
\right)
\]
where is basis order here is the same as for $U$.
The operator $S$ causes $\ket{010}$ to be exchanged with $\ket{100}$ and  $\ket{101}$ with $\ket{011}$, thus the operation is simply a SWAP between two of the qubits (the first two in our notation). In terms of the arrangement in Fig. 1(b), this would correspond to swapping $W$ and $X$. Of course, by suitably varying the composition of this operation, we can exchange any two selected qubits among the three adjacent to the barrier node. Thus these operations allows us to {\em freely permute} the arrangement of qubits within a network such as Fig. 2(b) or 2(c).

The operator $P$ simply associates a phase of $-1$ with the states $\ket{000}$ and $\ket{111}$. This is a kind of extension of the familiar two-qubit gate ``control-$\sigma_Z$''.  It is interesting to note that while the two-qubit operation of associating $-1$ with $\ket{00}$ and $\ket{11}$ can be composed from single qubit gates, this three-qubit analogue {\em cannot}. If, for example, we were to ensure that one of the qubits in the triplet was in state $\ket{0}$ (as we could simply by swapping a known $0$ into the triplet, using $S$) then we can create a control-NOT between the other two qubits simply via $XPX$ where $X\equiv \frac{1}{\sqrt 2}(\sigma_X+\sigma_Z)$ is the usual single-qubit Hadamard operation on the target qubit.

Thus $S$ and $P$, together with single-qubit gates, constitute a universal set of gates for QC. We conclude that the form of barrier gate illustrated in Fig. 1(b) does indeed suffice for gating interactions. 

\bigskip
\noindent {\bf General $J_{XY}\neq J_Z \neq0$ Models.}

We have seen that the limit of the pure $XY$ interaction ($J_Z=0$) is not suitable for this type of gate, but the limit of an isotropic Heisenberg interaction ($J_Z=J_{XY}$) is suitable. Away from these limits we have (with $d\equiv \delta/J_{XY}$ and $\Omega\equiv J_Z/J_{XY}$)
\[
\frac{ J_{XY}}{\pi}\tau_R=
\left\{
\begin{array}{c}
m(12+(2\Omega-d)^2)^{-1/2}\\
n(12+(2\Omega+d)^2)^{-1/2}\\
p(4+d^2)^{-1/2}\\
q(16+d^2)^{-1/2}
\end{array}
\right.
\]
In general, for an arbitrarily chosen $\Omega$ there is no solution. Are there any $\Omega$ values, other than the previously established $\Omega=1$, that will suffice? The answer is yes. For example, taking $d=0$ we find
\[
\frac{ J_{XY}}{\pi}\tau_R=
\left\{
\begin{array}{c}
n(12+4\Omega^2)^{-1/2}\\
p/2\\
q/4
\end{array}
\right.
\]
and we see that the condition that there be a solution is simply that $(12+4\Omega^2)^{-1/2}$ be a rational number. Thus there are an infinity of suitable $\Omega$ values, and indeed for any chosen $\Omega$ there will be an acceptable $\Omega$ arbitrarily close to it. Of course, to fully assess the practicality of using this scheme for such a value, one would need to generate the matrix corresponding to Eqn.[\ref{Umat}] and establish that it is entangling, find constructions for useful operators, etc.

We have seen that the three-qubit node can be used for QC, at least for a range of interactions that includes the important isotropic form. The natural question then becomes, can we extend the idea further to a four-qubit node?

\bigskip
\noindent{\bf The Four-Qubit Node.}

The analysis follows exactly as the three-qubit node, however with different conclusions. We consider a sub-network of nine spins with the topology and the initial state shown in Fig. 1(c), top. The full analysis is given in the Appendix. There we argue that, as with the three-qubit node, there is generally no advantage in considering cases where $\delta\neq 0$. With this simplification, the constraints for a revival are,

\begin{equation}
\frac{ J_{XY}}{\pi}\tau_R=
\left\{
\begin{array}{c}
m(8+\Omega^2)^{-\frac{1}{2}}\\
p(16+9\Omega^2)^{-\frac{1}{2}}\\
s(24+\Omega^2)^{-\frac{1}{2}}
\end{array}
\right.
\label{fourNodeEqs}
\end{equation}
where again $\Omega\equiv J_Z/J_{XY}$.

In general, one can choose an $\Omega$ to bring two of the three roots into a rational ratio - but the third will not be. There is only one apparent special case where all three can be satisfied, namely the $\Omega=1$ limit which corresponds to the important isotropic Heisenberg interaction. One might think, therefore, that at least in this one case the four-qubit node may function. However, inspecting the resulting conditions:
 \[
\frac{ J_{XY}}{\pi}\tau_R=m/3=p/5=s/5
\]
we find that the only solution is to set $\frac{ J_{XY}}{\pi}\tau_R$ equal to an integer, so the first revival is with $m=3$, $p=s=5$. When we enter this value of $\tau_R$ into the explicit expressions for state evolution, we find that the net effect of the entire process is simply the identity matrix. Thus, whereas for the three-qubit node we found that there was a useful revival at $\tau=\pi/(2J)$ prior to the second, identity revival at $\tau=\pi/J$, in the present case of the four-qubit node we find that there exists {\em only} the identity revival at $\tau=\pi/J$. Therefore we must conclude that it {\em is not possible} to implement a useful four-qubit gate in this way.

To summarize, we have examined two forms of multi-qubit gate based on a barrier scheme for QC which permits interactions to remain `always-on'. For the three-qubit node, we found that the isotropic Heisenberg interaction indeed yields an entangling gate operation on a time scale of $\pi/(2J)$, and confirmed that this operation can be used to construct more familiar SWAP and Control-NOT operations. Moreover, it seems possible that most, if not all, anisotropic Heisenberg interactions can be supported to a sufficient level of accuracy for QC. However, for the four-qubit node, we found that no useful operation can be performed. We have not ruled out the possibility that manipulating the qubit-bearing spins {\em during} the gate operation (by shifting Zeeman energies or by rapid flipping, for example) might generate a useful net operation, but this is beyond the scope of the basic barrier scheme. Possibly this is a direction for future work, as is the problem of finding optimal constructions for canonical gates via the three-qubit node. 

We conclude that the network topology shown in Fig. 2(d) is not feasible, but that the forms of network shown in Fig. 2(b) and (c) are feasible. These networks have a qubit storage efficiency that is $50\%$ higher than previously proposed 2D and 3D schemes. Moreover these networks are more efficient than a one-dimensional array, and may therefore represent the `best yet' embodiment for quantum computing with {\em always-on} interactions.

The author wishes to acknowledge support from a Royal Society URF, and from the Foresight LINK project  ``Nanoelectronics at the Quantum Edge''.

\newpage
\noindent {\large \bf Appendix: Analysis of the Four-Qubit Node.}

Consider a sub-network of nine spins with the topology and the initial state shown in Fig. 1(c), top. Now there are five barrier spins in the eigenstate $\ket{\uparrow}$ and four qubit bearing spins, the qubits being labelled $W$, $X$, $Y$ and $Z$.  As before we tune the Zeeman energy of the central barrier spin so that is comparable to the Zeeman energy of the qubit-bearing spins, and we assume for simplicity of analysis that the change is abrupt. By the same reasoning stated previously, the can neglect the outermost barriers except for the Zeeman shift they create. Then labelling  the four spins which initially bear qubits by the numbers $1$..$4$, and the middle spin by the letter $M$, the Hamiltonian for these five spins is exactly like Eqn.[\ref{hamil}] except that the index of course runs $j=1...4$. Complete representation of this Hamiltonian requires a matrix with 32 rows/columns, with the block-diagonal structure illustrated in Fig. 3(b). As before we will establish the eigenstructure of these subspaces, and then consider whether the `revivals' in each space can be made to coincide. The notation is analogous to that used before:
$\ket{\downarrow\downarrow\downarrow\downarrow}\ket{\uparrow}$, etc., where the left ket contains the states of the outer {\em four} spins, in clockwise order from the top left. Similarly we write the initial computation basis states as $\ket{0000}$, $\ket{0001}$...$\ket{1111}$, where the order is again clockwise from the top left, making the association exactly analogous to Eqn.[\ref{mapping}]. 

The two trivial subspaces are now associated with the states $\ket{\uparrow\uparrow\uparrow\uparrow}\ket{\uparrow}$ and $\ket{\downarrow\downarrow\downarrow\downarrow}\ket{\downarrow}$, which therefore remain eigenstates with energies $4J_Z\pm (5a+\delta)$.

The smallest non-trivial subspaces are now those with Hamiltonians $H_{4,1}$ and $H_{1,4}$. Again, these Hamiltonians will have the same structure due to symmetry. The form is:
\[
H_{4,1}= K\identity + 2J_{XY}\left( 
\begin{array}{ccccc}
0 & 0 & 0 & 0 & 1 \\ 
0 & 0 & 0 &  0 & 1 \\ 
0 & 0 & 0 &  0 & 1\\ 
0 & 0 & 0 &  0 & 1\\ 
1 & 1 & 1 &  1 & X
\end{array}
\right)
\]
For $H_{4,1}$ we have $K=(3a+\delta+2J_Z)$ and $X=X_{4,1}=-(\delta+3J_Z)/J_{XY}$, and the basis is 
$\{\ket{\downarrow\uparrow\uparrow\uparrow}\ket{\uparrow}$, $\ket{\uparrow\downarrow\uparrow\uparrow}\ket{\uparrow}$, 
$\ket{\uparrow\uparrow\downarrow\uparrow}\ket{\uparrow}$, 
$\ket{\uparrow\uparrow\uparrow\downarrow}\ket{\uparrow}$, 
$\ket{\uparrow\uparrow\uparrow\uparrow}\ket{\downarrow}\}$. 
For $H_{1,4}$ we have $K=(-3a-\delta+2J_Z)$ and $X=X_{4,1}=(\delta-3J_Z)/J_{XY}$, and the basis is 
$\{\ket{\uparrow\downarrow\downarrow\downarrow}\ket{\downarrow}$, $\ket{\downarrow\uparrow\downarrow\downarrow}\ket{\downarrow}$, 
$\ket{\downarrow\downarrow\uparrow\downarrow}\ket{\downarrow}$, 
$\ket{\downarrow\downarrow\downarrow\uparrow}\ket{\downarrow}$, 
$\ket{\downarrow\downarrow\downarrow\downarrow}\ket{\uparrow}\}$. 

We may write the (unnormalised) eigenvectors as:

\[
\Bigg\{ 
\ket{\rm{a}_1}=
\left( 
\begin{array}{c}
1  \\ 
-1  \\ 
0  \\ 
0  \\ 
0 
\end{array}
\right) ,\ \ \ \ \ket{\rm{a}_2}=
\left( 
\begin{array}{c}
0  \\ 
0  \\ 
1  \\ 
-1  \\ 
0 
\end{array}
\right) ,\ \ \ \ \ \ket{\rm{a}_3}=
\left( 
\begin{array}{c}
1  \\ 
1  \\ 
-1  \\ 
-1  \\ 
0 
\end{array}
\right) \Bigg\} \ \ \ \ \ \ \  \
\ket{V_-}=\left( 
\begin{array}{c}
1  \\ 
1  \\ 
1  \\ 
1  \\ 
V_- 
\end{array}
\right) \ \ \ \ \ \ \  \
\ket{V_+}=\left( 
\begin{array}{c}
1  \\ 
1  \\ 
1  \\ 
1  \\ 
V_+ 
\end{array}
\right)
\]
The vectors $\ket{V_\pm}$ have eigenvalues $E=V_\pm=\frac{1}{2}\left(X\pm(16+X^2)^{1/2}\right)$, while the bracketed triplet are degenerate with value $E=0$. Here $X$ is of course either $X_{4,1}$ or $X_{1,4}$ for the two respective subspaces. From these eigenvalues, the full energies are obtained as $K+ 2J_{XY}E$. By the same reasoning employed for the three-qubit node, we find one revival time constraint from each of these subspaces. From $H_{4,1}$ (where the initial state will be in the space spanned by the first four spin states, and at revival must return to this space):
\[
\tau_R=m\pi(16J_{XY}^2+(\delta+3J_Z)^2)^{-\frac{1}{2}}
\]
From $H_{1,4}$, where the initial state is $\ket{0000}=\ket{\downarrow\downarrow\downarrow\downarrow}\ket{\uparrow}$ and at revival must return to this state:
\[
\tau_R=n\pi(16J_{XY}^2+(\delta-3J_Z)^2)^{-\frac{1}{2}}.
\]

The remaining two subspaces, $H_{3,2}$ and  $H_{2,3}$, again have a common structure:
\[
H= K \identity+ 2J_{XY}\left( 
\begin{array}{cccccccccc}
0 & 0 & 0 & 0 & 0 & 0 & 0 & 0 & 1 & 1 \\ 
0 & 0 & 0 & 0 & 0 & 0 & 0 & 1 & 1 & 0 \\ 
0 & 0 & 0 & 0 & 0 & 0 & 1 & 1 & 0 & 0 \\ 
0 & 0 & 0 & 0 & 0 & 0 & 1 & 0 & 0 & 1 \\ 
0 & 0 & 0 & 0 & 0 & 0 & 1 & 0 & 1 & 0 \\ 
0 & 0 & 0 & 0 & 0 & 0 & 0 & 1 & 0 & 1 \\ 
0 & 0 & 1 & 1 & 1 & 0 & x & 0 & 0 & 0 \\ 
0 & 1 & 1 & 0 & 0 & 1 & 0 & x & 0 & 0 \\ 
1 & 1 & 0 & 0 & 1 & 0 & 0 & 0 & x & 0 \\ 
1 & 0 & 0 & 1 & 0 & 1 & 0 & 0 & 0 & x 
\end{array}
\right)
\]
For $H_{3,2}$ we have $K=(a+\delta)$ and $x=x_{3,2}=-(\delta+J_Z)/J_{XY}$, and the basis is 
$\ket{\downarrow\downarrow\uparrow\uparrow}\ket{\uparrow}$,
$\ket{\uparrow\downarrow\downarrow\uparrow}\ket{\uparrow}$,
$\ket{\uparrow\uparrow\downarrow\downarrow}\ket{\uparrow}$,
$\ket{\downarrow\uparrow\uparrow\downarrow}\ket{\uparrow}$,
$\ket{\uparrow\downarrow\uparrow\downarrow}\ket{\uparrow}$,
$\ket{\downarrow\uparrow\downarrow\uparrow}\ket{\uparrow}$ for the first six (the computational basis states), and 
$\ket{\uparrow\uparrow\uparrow\downarrow}\ket{\downarrow}$,
$\ket{\uparrow\uparrow\downarrow\uparrow}\ket{\downarrow}$,
$\ket{\uparrow\downarrow\uparrow\uparrow}\ket{\downarrow}$,
$\ket{\downarrow\uparrow\uparrow\uparrow}\ket{\downarrow}$ for the remaining four. For $H_{2,3}$ we have $K=-(a+\delta)$ and $x=x_{2,3}=(\delta-J_Z)/J_{XY}$, and the basis is of course the compliment of the $H_{3,2}$ basis.

The eigenstates include a degenerate pair $\{ \ket{{\rm asym}_1},\  \ket{{\rm asym}_2}\}$ with eigenvalue $E=0$ and two non-degenerate vectors $\ket{V_{24+}}$, $\ket{V_{24-}}$ with values $E=V_{24\pm}=\frac{1}{2}\left(x\pm(24+x^2)^{1/2}\right)$:
\[
\Bigg\{ \ket{{\rm asym}_1}=\left( 
\begin{array}{c}
1  \\ 
-1  \\ 
1  \\ 
-1  \\ 
0 \\ 
0 \\ 
0\\
0 \\
0 \\
0
\end{array}
\right) ,
\ \  \ket{{\rm asym}_2}=\left( 
\begin{array}{c}
1  \\ 
1  \\ 
1  \\ 
1  \\ 
-2  \\ 
-2  \\ 
0 \\
0 \\
0 \\
0 
\end{array}
\right)
\Bigg\}
\ \ \ \ \ \ \ 
 \ket{V_{24\pm}}=\left( 
\begin{array}{c}
2  \\ 
2  \\ 
2  \\ 
2  \\ 
2  \\ 
2  \\ 
V_{24\pm} \\
V_{24\pm} \\
V_{24\pm} \\
V_{24\pm} 
\end{array}
\right)
\]
 
The remaining eigenstates form two degenerate triplets, $\{\ket{V_{8-}a},\ket{V_{8-}b},\ket{V_{8-}c}\}$ and $\{\ket{V_{8+}a},\ket{V_{8+}b},\ket{V_{8+}c}\}$, with eigenvalues $E=V_{8\pm}=\frac{1}{2}\left(x\pm(8+x^2)^{1/2}\right)$.

\[
\Bigg\{ \ket{V_{8\pm}a}=\left( 
\begin{array}{c}
1  \\ 
1  \\ 
-1  \\ 
-1  \\ 
0  \\ 
0  \\ 
-V_{8\pm} \\
0 \\
V_{8\pm} \\
0 
\end{array}
\right) ,
\ \  \ket{V_{8\pm}b}=\left( 
\begin{array}{c}
1  \\ 
-1  \\ 
-1  \\ 
1  \\ 
0  \\ 
0  \\ 
0 \\
-V_{8\pm} \\
0 \\
V_{8\pm} 
\end{array}
\right) ,\ \  \ket{V_{8\pm}c}=\left( 
\begin{array}{c}
0  \\ 
0  \\ 
0  \\ 
0  \\ 
2  \\ 
-2  \\ 
V_{8\pm} \\
-V_{8\pm} \\
V_{8\pm} \\
-V_{8\pm}
\end{array}\right)
\Bigg\}
\]

all eigenvalues corresponding of course to total energies $\delta-J_Z+2J_{XY}E$.

We can now consider the motion of states initially in the computational subspaces. First consider the $H_{2,3}$ space: here the computational subspace is $\{\ket{0001}$, $\ket{0010}$, $\ket{0100}$, $\ket{1000}\}$. By analysing the development of a state initially in this space, exactly as was done for the three-qubit node, we find two constraints on the revival time:
\[
\tau_R=
\left\{
\begin{array}{c}
m\pi(8J_{XY}^2+(\delta-J_Z)^2)^{-\frac{1}{2}}\\
n\pi(24J_{XY}^2+(\delta-J_Z)^2)^{-\frac{1}{2}}
\end{array}
\right.
\]
Similarly, from the $H_{3,2}$ space containing the six computational basis states $\{\ket{1100}$, $\ket{0110}$, $\ket{0011}$, $\ket{1001}$, $\ket{1010}$, $\ket{0101}\}$, we find that the condition for a state, initially within that space, to return to it is:
\[
\tau_R=
\left\{
\begin{array}{c}
m\pi(8J_{XY}^2+(\delta+J_Z)^2)^{-\frac{1}{2}}\\
n\pi(24J_{XY}^2+(\delta+J_Z)^2)^{-\frac{1}{2}}
\end{array}
\right.
\]
Combining these conditions with those from the smaller $H_{4,1}$ \& $H_{1,4}$ spaces, we have our complete set of revival time conditions:
\[
\frac{ J_{XY}}{\pi}\tau_R=
\left\{
\begin{array}{c}
m(8+(\Omega+d)^2)^{-\frac{1}{2}}\\
n(8+(\Omega-d)^2)^{-\frac{1}{2}}\\
p(16+(3\Omega+d)^2)^{-\frac{1}{2}}\\
q(16+(3\Omega-d)^2)^{-\frac{1}{2}}\\
r(24+(\Omega+d)^2)^{-\frac{1}{2}}\\
s(24+(\Omega-d))^2)^{-\frac{1}{2}}
\end{array}
\right.
\]
As before, we observe that there will exist some set of integers $m..s$ which indeed mutually satisfy these conditions if, and only if, all $()^{-\frac{1}{2}}$ terms are be related to one another in rational ratios. We note that there is no advantage in using $d\neq 0$ since this doubles the number of conditions while only providing one variable. Therefore using $d=0$ we find
\[
\frac{ J_{XY}}{\pi}\tau_R=
\left\{
\begin{array}{c}
m(8+\Omega^2)^{-\frac{1}{2}}\\
p(16+9\Omega^2)^{-\frac{1}{2}}\\
s(24+\Omega^2)^{-\frac{1}{2}}
\end{array}
\right.
\]
as quoted in the body of the paper, Eqn.[\ref{fourNodeEqs}].


\begin{references}
\bibitem{notNMR} NMR systems are a partial exception to this rule: there the interaction is of the diagonal Ising form and is weak compared to the rate at which single spins can be rotated; the combination of these two special factors allows the `always on' interaction to be effectively negated by methods such as periodic spin flipping.
\bibitem{1Dlimits} A one-dimensional architecture that relies on the nearest-neighbor interactions must pay a heavy overhead in terms of the proportion of operations that are simple SWAPs, needed merely to permute qubits. This is due to the average separation between a given qubit pair being $\propto N$ in a 1D array, versus $N^{1/2}$ or $N^{1/3}$ in two- or three-dimensional networks. There are consequences for fundamental considerations such as the tolerable error rate etc.
\bibitem{composites} The following argument will allow the reader to verify our claim that $S$ and $P$ can be formed. These constructions are in no sense optimal - finding the efficient compositions for specific gates is beyond the scope of the present paper. Let $R_Z(\alpha,\beta,\gamma)$ be the operator that applies a single qubit Z-rotation $\cos \alpha \identity + i\sin\alpha\sigma_Z$ to the first qubit, while independently rotating the second and third by $\beta$ and $\gamma$ respectively. Introduce special cases $R_\pm=R_Z(x,-x,0)$ with $x=\pm(\pi/4+\arcsin(1/8)/2)$. Let $U_\pm=U.R_\pm.U$. Then defining $W\equiv R(-\pi/2+y,-\pi/2-y,-\pi/2).U_+.R(0,0,\pi/2).U_-$ where $y=\arctan(-\sqrt{7}/3)/2$, we notice $W=P.S$ i.e. it is composed of the two operators we are interested in. Further, we find that $T_- R(\pi/4,\pi/4,0).W.T_{YZ}.W.T_{YZ}.T_+=P$, where single-qubit gates $T_{YZ}=\frac{1}{\sqrt{2}}(\sigma_Y+\sigma_Z)$ and $T_\pm=\frac{1}{\sqrt{2}}(\identity\pm i\sigma_Y)$ all act on the third qubit. Given $P$ we know we must also be able to obtain $S$ since $S=P.W$.
\end{references}
\end{document}